\documentclass{jfm}
\usepackage{graphicx}
\usepackage{amssymb}
\usepackage{amsmath}
\usepackage{bm}
\usepackage{lineno}

\DeclareMathOperator\erf{erf}

\shorttitle{Edge vortex of a revolving plate}
\shortauthor{D. Chen, D. Kolomenskiy and H. Liu}

\title{Closed-form solution for the edge vortex of a revolving plate}

\author{
  Di Chen\aff{1, 2, 3}
  Dmitry Kolomenskiy\aff{2, 4}
  \corresp{\email{dkolom@gmail.com}},
  \and
  Hao Liu\aff{2, 3}
  \corresp{\email{hliu@faculty.chiba-u.jp}}
  }

\affiliation
  {
  \aff{1}
  School of Naval Architecture, Ocean and Civil Engineering, Shanghai-Jiao Tong University, Shanghai, People's Republic of China
  \aff{2}
  Graduate School of Engineering, Chiba University, Chiba, Japan
  \aff{3}
  Shanghai-Jiao Tong University and Chiba University International Cooperative Research Center (SJTU-CU ICRC), Shanghai, People's Republic of China
  \aff{4}
  Center for Earth Information Science and Technology (CEIST),\\ Japan Agency for Marine-Earth Science and Technology (JAMSTEC), Yokohama, Japan
  }

\begin{document}

\maketitle

%\linenumbers

\begin{abstract}
Flapping and revolving wings can produce attached leading edge vortices (LEVs) when the angle of attack is large.
In this work, a low order model is proposed for
the edge vortices that develop on a revolving plate at 90 degrees angle of attack,
which is the simplest limiting case, yet showing remarkable similarity with the generally known LEVs.
The problem is solved analytically, providing short closed-form expressions for the circulation
and the position of the vortex. A good agreement with the numerical solution of the Navier--Stokes equations
suggests that, for the conditions examined, the vorticity production at the sharp edge and its subsequent three-dimensional transport
are the main effects that shape the edge vortex.
\end{abstract}

\begin{keywords}
%Authors should not enter keywords on the manuscript, as these must be chosen by the author during the online submission process and will then be added during the typesetting process (see http://journals.cambridge.org/data/\linebreak[3]relatedlink/jfm-\linebreak[3]keywords.pdf for the full list)
\end{keywords}

\section{Introduction}

Separated flows over flapping or revolving flat plates
have gained attention over the past decades in the context of
animal locomotion and insect flight in particular.
Wings of insects have sharp edges that
generate leading edge vortices (LEVs)
responsible for the high lift coefficient
at large angles of attack \cite[]{Ellington_etal_1996_nature,Liu_etal_1998_jeb}.
%To some extent, the lift enhancement is explained by
%two-dimensional unsteady mechanisms \cite{Wang_2005_arfm},
The aerodynamics of flapping wings combines multiple lift-enhancement mechanisms.
However, experiments with unilaterally rotating wings by \cite{Maxworthy_1979_jfm,Usherwood_Ellington_2002b_jeb,Lentink_Dickinson_2009b_jeb}
have shown similar lift enhancement and LEV structures as flapping wings in the middle of downstroke and upstroke,
and it has been recognized that the three-dimensional character of the flow
is important therewith.

The shape of an LEV on a flapping or a revolving wing is approximately conical,
it expands with the distance from the axis of revolution until it separates at some spanwise location
where its size becomes commensurate with the wing local chord length \cite[]{Kruyt_etal_2015_interface}.
The conical vortex leaves a triangular low-pressure footprint on the upper surface near the leading edge of the wing, thus
producing net lift.
This effect persists over a wide range of flow regimes, despite transitions from a
steady laminar diffuse LEV when the Reynolds number is of order $Re=100$ to
a more compact conical vortex core at $Re=1000$,
then to a turbulent LEV at $Re$ of order $10000$ \cite[]{Usherwood_Ellington_2002b_jeb,Garmann_2013_pof}.
It is likely that the spanwise flow from the wing root to the tip
is critical for shaping up a steady LEV by removing the vorticity spanwise
and depositing it into a trailing vortex \cite[]{Maxworthy_1979_jfm,Liu_etal_1998_jeb}.
Alternative explanations based on PIV measurements include the effect of downward flow induced by tip vortices \cite[]{Birch_Dickinson_2001_nature} and
vorticity annihilation due to interaction between the LEV and the opposite-sign layer on the wing \cite[]{Wojcik_Buchholz_2014_jfm}.

As compared with the substantial amount of recent experimental and numerical work \cite[for a review see, e.g.,][]{Limacher_etal_2016_jfm},
only few analytical or low-order models have been proposed to understand the LEV dynamics of revolving wings.
\cite{Maxworthy_2007_jfm} derived an estimate for the spanwise velocity.
\cite{Limacher_etal_2016_jfm} studied the role of Coriolis accelerations.
However, no estimate has been proposed for such an important quantity as the circulation.
In \S\ref{sec:math_formulation} of the present paper, we derive closed-form expressions in elementary functions for the circulation and the position of the edge vortex.
For simplicity, we restrict our attention to a rectangular plate at $90^\circ$ angle of attack.
The edge vortex of this plate is nominally similar to the LEV of a plate at any large angle of attack,
with the main difference of the downwash vanishing at the angle of $90^\circ$.
%use the Brown--Michael point vortex approximation \cite[]{Brown_Michael_1954_aiaa} in the present contribution.
%This problem is in many respects similar to a more commonly studied case of the 45 deg angle of attack,
%but avoids complications due to downwash.
Good agreement with the numerical solution of the incompressible Navier--Stokes equations
shown in \S\ref{sec:discussion} suggests that, for the conditions examined, the vorticity production at the edge and its subsequent transport downstream and spanwise
are likely the main effects that explain the circulation and the location of the vortices observed in the numerical simulations.
Implications of these findings and perspectives for future improvement of the model are discussed in \S\ref{sec:conclusions}.

\section{Mathematical formulation of the edge vortex model}\label{sec:math_formulation}

The wing considered in this study is a flat plate with sharp edges. It is set at a
constant angle of attack $90^\circ$ and revolves with a constant
angular velocity $\Omega$ about the vertical axis, as shown in figure~\ref{fig:3d_scheme}(\textit{a}).
For simplicity of the analysis, we suppose that the planar shape of the plate is rectangular
with length $R$ and chord $c$,
and that the axis of revolution passes through the root edge.
Due to the top-bottom symmetry of the setup,
we only focus on the flow above the symmetry plane,
and the ``edge vortex" refers to the vortex near the top edge of the plate,
unless we explicitly state the opposite.

\subsection{Line vortex model}

Earlier studies have revealed a nominally conical shape of the edge vortex,
which expands from the root towards the tip of the plate.
In a reference frame revolving with the plate, the flow
is essentially in the azimuthal direction
and in the spanwise direction from the root to the tip.
Therefore, the flow over the nearest sharp edge
is likely to be the key factor that determines how the edge vortex develops
over the proximal portion of the plate.
The influence of the finite span of the plate
only becomes strong near its distal part, and this effect is neglected
in the present analysis.
The effect of the finite chord length is taken into account approximately by
using potential flow asymptotics for the velocity.

\begin{figure}
\begin{center}
\includegraphics[scale=0.79]{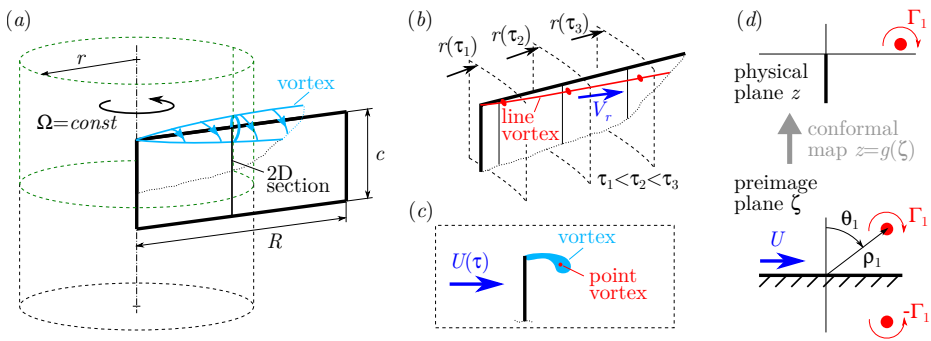}
\caption{\label{fig:3d_scheme} (\textit{a}) Drawing of a revolving plate highlighting the edge vortex domain considered in our analysis. (\textit{b}) Line vortex model and the radial position of a Lagrangian vortex element at consequent time instants $\tau_1$, $\tau_2$ and $\tau_3$. (\textit{c}) Two-dimensional point vortex approximation of the flow. (\textit{d}) The flow domain in the physical plane $z$ and in the preimage plane $\zeta$.}
\end{center}
\end{figure}

Thus, the viscous flow in a small neighborhood around the edge
is dominated by the separation that produces vorticity.
In two-dimensional flows, or if the plate is in pure translation, the vorticity accumulates in the near wake region
until it sheds as a separated vortex.
The flow topology changes dramatically due to the presence of the spanwise flow
that removes the vorticity from the edge vortex and deposits it into a trailing vortex
when the wing revolves \cite[]{Maxworthy_1979_jfm,Liu_etal_1998_jeb,Lentink_Dickinson_2009b_jeb}.
Hence, among all of the effects that have any influence on the edge vortex properties,
we postulate that two phenomena are of utter importance: (i) vorticity production and
(ii) three-dimensional transport of the vorticity. Using approximate models of these two phenomena,
we derive the desired estimates for the edge vortex position and circulation.

The next important step is to approximate the diffuse vortex core by a thin vortex line
that originates from the root and extends toward the tip of the plate, see figure~\ref{fig:3d_scheme}(\textit{b}).
An element $\mathrm{d} r$ of that line vortex at a distance $r$ from the axis of revolution
substitutes for the radial vorticity in the fluid contained between two virtual cylinders
of radii $r$ and $r+\mathrm{d} r$. Our model neglects the vorticity components in the directions other than the radial.
The error is estimated \textit{a posteriori} in Appendix~\ref{sec:error_local_approx}.
We follow the path of a selected Lagrangian element of the line vortex
as its distance from the axis of revolution $r(\tau)$ increases in time $\tau$
due to the spanwise advection, and
use the Brown--Michael vortex to estimate the vorticity produced
at any $r$.
The Lagrangian vortex particle moves spanwise with the velocity $V_r$ such that
$\mathrm{d} r / \mathrm{d} \tau = V_r(r)$. We postulate that it is related to the inflow velocity
$U(r)$ as
\begin{equation}
V_r = K_{sp} U, \quad \mathrm{where} \quad U = \Omega r,
\label{eq:vsp_original}
\end{equation}
and earlier research by \cite{Maxworthy_2007_jfm} and \cite{Limacher_etal_2016_jfm}, as well as our numerical simulations suggest that it is adequate to assume $K_{sp}=const$.
After integration we obtain
\begin{equation}
r(\tau) = r_0 e^{K_{sp} \Omega \tau} \quad \mathrm{and} \quad U(\tau) = \Omega r_0 e^{K_{sp} \Omega \tau},
\label{eq:r_lagrangian}
\end{equation}
where $r_0$ is an integration constant.
We thus reduce the three-dimensional steady problem to a two-dimensional unsteady problem
of vortex dynamics on a cylinder of radius $r(\tau)$, and substitute it with
a Brown--Michael model of the flow over a sharp edge, see figure~\ref{fig:3d_scheme}(\textit{c}).
All three-dimensional effects other than the spanwise advection are neglected at this point.
%Implications of that approximation will be discussed later.
% Note that, in the neighborhood of the leading edge it is justified to neglect the curvature.

\subsection{Solution of the Brown--Michael model}\label{sec:steady_theory}

The Brown--Michael model for the flow past a semi-infinite plate perpendicular to the free stream
was solved by \cite{Cortelezzi_1995_pof}. We briefly repeat the derivation with only a slight modification
of explicitly entering the chord length $c$ in the equation, for the ease of comparison with numerical simulations. % later in this paper.
%of explicitly entering the chord length $c$ in the equation, for the ease of comparison with numerical simulations. % later in this paper.
% More details are given in the Supplementary Material.

The physical flow domain is an infinite space with a vertical plate immersed in the fluid.
The origin of the coordinate system $z=0$ is at the top edge of the plate.
Using a conformal mapping
\begin{equation}
z = g(\zeta), \quad \textrm{where} ~~ g(\zeta) = - \mathrm{i} \zeta^2 / c,
\label{eq:mapping}
\end{equation}
%with $\zeta$ being the preimage coordinate,
the leading-order term of the flow near the edge is mapped
on the complex half-plane, as shown in figure~\ref{fig:3d_scheme}(\textit{d}).
The point vortex has strength $\Gamma_1$
and position $z_1$ that vary in time $\tau$, and in the following we derive explicit solutions
for these two quantities.
Since it is obvious that the flow generates a clockwise vortex,
we follow the convention of \cite{Cortelezzi_1995_pof} that assumes that clockwise circulation is positive.
The complex potential of the flow is equal to
\begin{equation}
W(\zeta,\tau) = U(\tau) \zeta - \frac{\Gamma_1(\tau)}{2 \pi \mathrm{i}} \ln \frac{\zeta - \zeta_1(\tau)}{\zeta - \zeta_1^*(\tau)}.
\label{eq:complex_pot}
\end{equation}
The Kutta condition is satisfied if $\partial W / \partial \zeta = 0$ at $\zeta=0$, which determines the circulation
\begin{equation}
\Gamma_1(\tau) = 2 \pi \mathrm{i} \frac{\zeta_1(\tau) \zeta_1^*(\tau)}{\zeta_1(\tau)-\zeta_1^*(\tau)} U(\tau).
\label{eq:gamma1_def}
\end{equation}
The unknown position of the vortex $\zeta_1$ is obtained from the Brown--Michael equation
\begin{equation}
\frac{\mathrm{d} z_1^*}{\mathrm{d} \tau} + z_1^* \frac{1}{\Gamma_1} \frac{\mathrm{d}\Gamma_1}{\mathrm{d}\tau} = \tilde{u}^*
\label{eq:bm_motion}
\end{equation}
with $z^*_1(0) = 0$ as the initial condition.
The de-singularized complex conjugate velocity of the point vortex in the physical plane is equal to
\begin{equation}
\tilde{u}^* = \frac{\mathrm{i} c}{2 \zeta_1} \left\{ U(\tau) - \frac{\mathrm{i} \Gamma_1(\tau)}{2 \pi} \frac{1}{\zeta_1(\tau) - \zeta_1^*(\tau)} - \frac{\mathrm{i} \Gamma_1(\tau)}{4 \pi} \frac{g''(\zeta_1(\tau))}{g'(\zeta_1(\tau))} \right\}.
\label{eq:utilde}
\end{equation}
After substituting (\ref{eq:utilde}), (\ref{eq:gamma1_def}) and (\ref{eq:mapping}) into (\ref{eq:bm_motion}),
we obtain an ordinary differential equation for $\zeta_1$,
\begin{equation}
\left( 2 \mathrm{i} \zeta_1^* + \frac{\mathrm{i} \zeta_1 \zeta_1^*}{\zeta_1 - \zeta_1^*} \right) \frac{\mathrm{d} \zeta_1^*}{\mathrm{d} \tau} -
\frac{\mathrm{i} {\zeta_1^*}^3}{\zeta_1 (\zeta_1-\zeta_1^*)} \frac{\mathrm{d} \zeta_1}{\mathrm{d} \tau} + \frac{\mathrm{i} {\zeta_1^*}^2}{U} \frac{\mathrm{d} U}{\mathrm{d} t} =
\frac{\mathrm{i} c^2}{2 \zeta_1} \left( U - \frac{\mathrm{i} \Gamma_1}{2 \pi (\zeta_1-\zeta_1^*)} - \frac{\mathrm{i} \Gamma_1}{4 \pi \zeta_1} \right)
\label{eq:ode_zeta1}
\end{equation}
with the initial condition $\zeta_1(0)=0$.
In the polar coordinates $\rho_1$ and $\theta_1$ such that $\zeta_1=\rho_1 e^{\mathrm{i}(\pi/2-\theta_1)}$, equation (\ref{eq:ode_zeta1})
is equivalent to a system of two equations,
\begin{equation}
\begin{split}
\frac{\mathrm{d} \rho_1}{\mathrm{d} \tau}   & = \frac{c^2 U}{12 \rho_1^2} \sin \theta_1 - \frac{\rho_1}{3U} \frac{\mathrm{d} U}{\mathrm{d} \tau}, \\
\frac{\mathrm{d} \theta_1}{\mathrm{d} \tau} & = \frac{c^2 U}{8 \rho_1^3} \frac{\cos 2 \theta_1}{\cos \theta_1}
\end{split}
\label{eq:polar_eqns}
\end{equation}
with the initial conditions
\begin{equation}
\begin{split}
\rho_1(0) & = 0, \\
\theta_1(0) & = \theta_0, \quad \theta_0 \in ]-\pi/2,\pi/2[.
\end{split}
\end{equation}
After the change of variables
$\Upsilon = U \rho_1^3 / c^2$, $\Theta = \sin \theta_1$ and
\begin{equation}
\tilde{\tau} = \int_{-\infty}^\tau U^2(\tau') \mathrm{d} \tau' = \frac{\Omega r^2}{2 K_{sp}}
\label{eq:tau_def}
\end{equation}
that makes use of (\ref{eq:r_lagrangian}),
equations (\ref{eq:polar_eqns}) transform into
\begin{equation}
\frac{\mathrm{d} \Upsilon}{\mathrm{d} \tilde{\tau}} = \frac{\Theta}{4}, \quad
\frac{\mathrm{d} \Theta}{\mathrm{d} \tilde{\tau}} = \frac{1 -2 \Theta^2}{8 \Upsilon}
\end{equation}
with $\Upsilon(0) = 0$ and $\Theta(0) = \Theta_0 \in ]-1,1[$.
Combining the two equations, we obtain an equation of the second order,
\begin{equation}
\frac{\mathrm{d}^2 (\Upsilon^2)}{\mathrm{d} \tilde{\tau}^2} = \frac{1}{16},
\end{equation}
that has two branches of the solution $\Upsilon = \pm \tilde{\tau} / \sqrt{32}$ satisfying the desired boundary condition,
and we choose the `+' sign which is the physically relevant one.
We therefore find $\Theta = \sqrt{2}/2$ and
\begin{equation}
\rho_1 = \left( \frac{c^2 \tilde{\tau}}{2^{5/2} U} \right)^{1/3}, \quad \theta_1 = \frac{\pi}{4}.
\end{equation}
Noting that $\tilde{\tau}/U=r/2K_{sp}$ and mapping the solution to the physical plane using (\ref{eq:mapping}),
we obtain the position of the vortex as a function of distance $r$ from the axis of revolution,
\begin{equation}
\frac{z_1}{c} = \frac{1}{2^{7/3}K_{sp}^{2/3}} \left( \frac{r}{c} \right)^{2/3}.
\label{eq:pos_90deg}
\end{equation}
Even though $z_1$ is a complex number by definition, the imaginary part of (\ref{eq:pos_90deg}) is zero.
The circulation is obtained from (\ref{eq:gamma1_def}). In polar coordinates
it simplifies to $\Gamma_1 = \pi \rho_1 U / \cos{\theta_1}$, yielding
\begin{equation}
\frac{\Gamma_1}{\Omega c^2} = \frac{\pi}{(4K_{sp})^{1/3}} \left( \frac{r}{c} \right)^{4/3}.
\label{eq:circ_90deg}
\end{equation}
Equations (\ref{eq:pos_90deg}) and (\ref{eq:circ_90deg}) are the main results of this paper.

\subsection{Numerical solution of the Navier--Stokes equations}

For validation of the theoretical model, we employ established
tools of the computational fluid dynamics (CFD).
The incompressible three-dimensional Navier--Strokes equations are solved using a commercial finite-volume code ANSYS CFX 14.5.
We consider a plate with the chord length $c = 1$~mm
and uniform thickness $0.02c$.
The distance from the axis of revolution to the tip is equal to $R=6c$
in all numerical simulations except one which is described separately in the end of \S\ref{sec:comparison}.
The plate is immersed in a spherical inner domain of radius $10c$,
and both rotate around the vertical axis with
the angular velocity that gradually increases with time $t$ as $0.5 \Omega (1-\cos(\pi t/t_{ac}))$
until it becomes equal to $\Omega$, then remains constant during all $t > t_{ac}$ \cite[cf.][]{Harbig_etal_2013_jfm}.
The acceleration time $t_{ac}$ is equal to $0.0835T$,
where $T = 2 \pi / \Omega$.
The outer stationary domain is a cuboid with its top,
bottom and side far-field boundaries located at, respectively, $120c$, $120c$ and $80c$ away from the center of the inner spherical domain.
The domains are discretized with hexahedron meshes of high
quality, with the minimum grid spacing adjacent to the wall surface $\delta_{min}=0.1\sqrt{\nu c / \Omega R}$.
The General Grid Interface (GGI) technique is applied to connect the two domains
in a Multiple Frame of Reference (MFR), and a moving grid method is utilized in the inner domain.
The grids have about 2.54 million cells in the simulations with $\Omega$ equal to $130$, $260$ and $520~\mathrm{s}^{-1}$.
The case of $\Omega=1300~\mathrm{s}^{-1}$ requires 4.61 million cells to ensure the
same accuracy.
The Courant number is approximately equal to 1 in all simulations.
The kinematic viscosity of the fluid is equal to $\nu = 1.56 \cdot 10^{-5}$~m$^2$/s.
The near field of the plate reaches a seemingly steady state by $t=0.8T$,
therefore, instantaneous flow fields at that time instant are used for the
comparison with the theoretical estimates.

\section{Discussion}\label{sec:discussion}

\subsection{Comparison between the analytical and the numerical solutions}\label{sec:comparison}

The output of our model is the circulation (\ref{eq:circ_90deg}) and
the position (\ref{eq:pos_90deg}) of the vortex as functions of $r/c$.
These are well defined quantities for a line vortex,
but there exist many alternative definitions of a vortex when it has a diffuse core.
For an objective comparison between the theoretical estimates and
the CFD results, let us not restrict our attention to the vorticity in the core.
Instead, since the flow over the plate at $90^\circ$ is symmetric, let us consider
the circulation $\Gamma_\Sigma(r)$ obtained by integrating the radial vorticity component $\omega_r$
over the entire half-cylinder of radius $r$ above the symmetry plane shown with green dashed lines in figure~\ref{fig:3d_scheme}(\textit{a}).
When using $\omega_r$ obtained from the CFD,
the vertical extent of the domain is truncated at $L_y/2=10c$, yielding
\begin{equation}
%\Gamma_\Sigma^{CFD} = \int_{0}^{2 \pi} \left\{ \int_{0}^{L_y/2} \omega_r \mathrm{d} y \right\} r \mathrm{d} \phi.
\Gamma_\Sigma^{CFD} = \int_{0}^{2 \pi} \int_{0}^{L_y/2} \omega_r r \mathrm{d} y \mathrm{d} \phi,
\label{eq:circulation_cfd_90deg}
\end{equation}
where $y$ is the vertical coordinate and $\phi$ is the azimuthal coordinate.

\begin{figure}
\begin{center}
\includegraphics[scale=0.905]{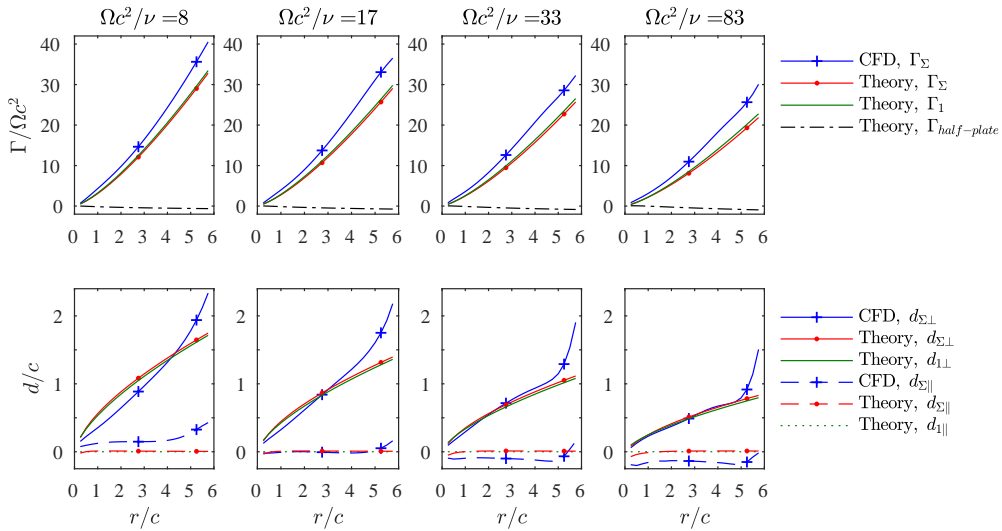}
\caption{\label{fig:gam_dist_aoa90} Comparison between the theoretical estimates and the CFD results
for the circulation $\Gamma_\Sigma$ over a cylinder surface of radius $r$ (top row), and
the distances between the edge and the vorticity centroids in
the directions parallel and perpendicular to the plate, $d_\parallel$ and $d_\perp$, respectively (bottom row). Estimates
for the point vortex circulation $\Gamma_1$, components of its distance to the edge $d_{1\perp}$, $d_{1\parallel}$, and
the half-plate circulation $\Gamma_{half-plate}$ are added for reference. All quantities are normalized.}
\end{center}
\end{figure}

On the other hand, in the theoretical model, the line vortex substitutes for all vorticity in the
entire domain with the exception of the boundary layers on the plate.
The boundary layer vorticity is represented by the ``bound'' circulation
along a contour that intersects with the plate but does not encompass the point vortex in the physical fluid domain.
The bound circulation of a half-plate is estimated using
the values of $W$ given by (\ref{eq:complex_pot}) at $c/2$ distance from the edge, on the pressure and on the suction side of the plate (see Appendix~\ref{sec:half-plate} for the derivation),
resulting in
\begin{equation}
\frac{\Gamma_{half-plate}}{\Omega c^2} = \sqrt{2} \frac{r}{c} - \frac{\Gamma_1}{\Omega c^2} \left( \frac{1}{2} + \frac{1}{\pi} \arctan{\frac{1-2z_1/c}{2\sqrt{z_1/c}}} \right),
\label{eq:circ_half_plate}
\end{equation}
where $z_1$ is a real number, as given by (\ref{eq:pos_90deg}).
The theoretical estimate for $\Gamma_\Sigma(r)$ is therefore
\begin{equation}
\Gamma_{\Sigma}^{Theory} = \Gamma_1 + \Gamma_{half-plate},
\end{equation}
with the two components on the right-hand side evaluated using
(\ref{eq:circ_90deg}) and (\ref{eq:circ_half_plate}), respectively. %, and $K_{sp}$ evaluated using (\ref{eq:ksp_estimate_Re}).

The top row panels in figure~\ref{fig:gam_dist_aoa90} present a comparison between $\Gamma_{\Sigma}^{Theory}$ and $\Gamma_{\Sigma}^{CFD}$
at different flow regimes characterized by the root-based Reynolds number $\Omega c^2 / \nu$ in the range between 8 and 83. The equivalent Reynolds
number based on the wing-tip velocity and the chord length $Re=\Omega R c / \nu$ is in the range $Re=50...500$.
All quantities are normalized.
The agreement between the theoretical and the numerical results is good in all cases.
The shape of the profiles makes the theoretical 4/3 power law apparent, while the good pointwise agreement
is ensured by substituting $K_{sp}$ with a fit
\begin{equation}
\tilde{K}_{sp} = 0.078 \sqrt{\Omega c^2 / \nu}
\label{eq:ksp_estimate_Re}
\end{equation}
that minimizes the root mean square error, as discussed in the next section.
Note that, even if $\tilde{K}_{sp}$ only depends on $\Omega c^2 / \nu$,
the dimensionless circulation (\ref{eq:circ_90deg}) and position (\ref{eq:pos_90deg}) depend on $r/c$ as well.
It is straightforward, however, to derive a normalization that yields normalized $z_1$ and $\Gamma_1$
being functions of the root-based Reynolds number only:
$z_1 / (r^2 c)^{1/3} = 1.087 (\Omega c^2 / \nu)^{-1/3}$ and $\Gamma_1 / \Omega (r^2 c)^{2/3} = 4.632 (\Omega c^2 / \nu)^{-1/6}$.
Similar expressions can be written in terms of the local spanwise Reynolds number $\Omega r c / \nu$.

\begin{figure}
\begin{center}
\includegraphics[scale=0.795]{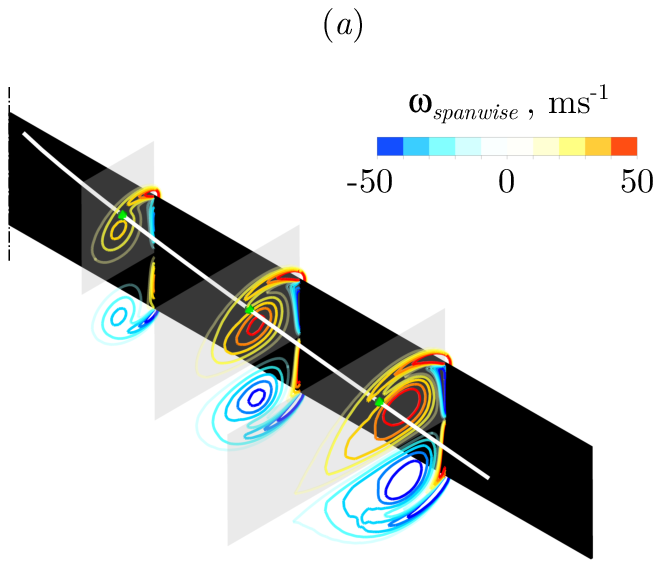}\quad\quad
\includegraphics[scale=0.880]{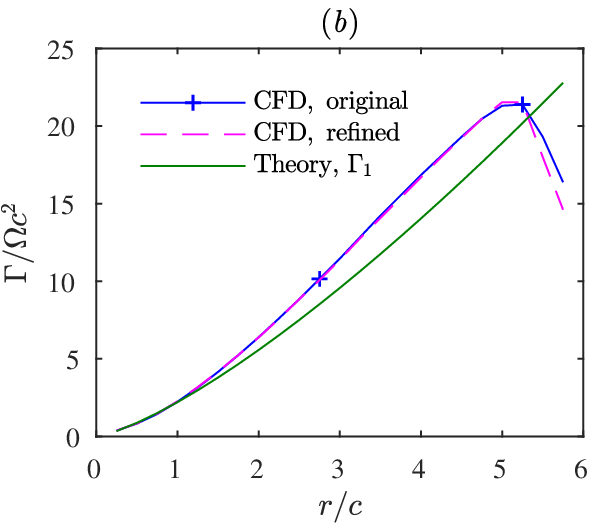}
\caption{\label{fig:cmp_re500} Flow over the plate at $\Omega c^2/\nu = 83$.
(\textit{a}) Vorticity isocontours superposed with the theoretical position of the line vortex $z_1$.
Vorticity magnitude scale is in milliseconds$^{-1}$.
Gray shaded rectangles
highlight the integration domain used for calculation of the vortex circulation in this example.
(\textit{b}) Normalized circulation of the vortex as a function of the
normalized spanwise distance.}
\end{center}
\end{figure}

%Even near the wing tip the agreement is good.
As $Re$ increases, $\Gamma_{\Sigma}$ becomes smaller.
This is related to the vortex becoming nearer to the edge, as shown in
the bottom row panels in figure~\ref{fig:gam_dist_aoa90}, in terms of the
components of the distance between the edge of the plate $z_{edge}$ and the vorticity central line $z_\Sigma$ in the directions perpendicular
and parallel to the plate, $d_\perp=\Re(z_\Sigma-z_{edge})$ and $d_\parallel=\Im(z_\Sigma-z_{edge})$, respectively.
The vorticity central line $z_\Sigma(r)$ in the CFD is calculated as
\begin{equation}
%z_\Sigma^{CFD} = \frac{1}{\Gamma_\Sigma^{CFD}} \int_{0}^{2 \pi} \left( \int_{0}^{L_y/2} (r \phi + \mathrm{i} y) \omega_r \mathrm{d} y \right) r \mathrm{d} \phi.
z_\Sigma^{CFD} = \frac{1}{\Gamma_\Sigma^{CFD}} \int_{0}^{2 \pi} \int_{0}^{L_y/2} (r \phi + \mathrm{i} y) \omega_r r \mathrm{d} y \mathrm{d} \phi.
\label{eq:z_cfd_90deg}
\end{equation}
This definition is equally suitable for flows at any $Re$, including those cases when it is difficult to identify the vortex core.
Its counterpart in the line vortex model is
\begin{equation}
z_\Sigma^{Theory} = \frac{z_1~\Gamma_1 ~+~ z_{half-plate}~\Gamma_{half-plate}}{\Gamma_1 + \Gamma_{half-plate}},
\end{equation}
where $z_{half-plate}$ is
calculated using the distribution of bound vorticity over the plate, as explained in Appendix~\ref{sec:half-plate}.
The agreement between the theoretical estimate and the results of the numerical
simulation is the best over the inner-central part of the plate.
%The discrepancy increases as $r$ approaches the tip of the plate.
%It may be attributed the Coriolis acceleration which is not taken into account in the theoretical model.
%However, this effect is secondary as the discrepancy does not exceed 30\% for all $r/c<4.5$.
When $r/c>5$, the wing tip effects become dominant and the vorticity spreads far behind the plate in the CFD results.
This effect is beyond the limitations of our theoretical model of the edge vortex that neglects aerodynamic interactions
with the wing tip.

When $Re$ is sufficiently large,
the edge vortex has a distinguishable core of large
axial vorticity. Let us compare its properties
with the line vortex model estimate at $\Omega c^2/\nu = 83$.
In PIV experiments as well as in numerical simulations,
the circulation is usually calculated by summing up
the spanwise vorticity contained in flat rectangular windows, cf. \cite{Carr_etal_2015_jfm}.
Therefore, in this example, we also use flat windows of hight $c$ and width $0.5 r$, shown
as gray shaded areas in figure~\ref{fig:cmp_re500}(\textit{a}).
Sectional isolines of the vorticity component perpendicular to the integration planes
reveal the vortex core. The white line
superposed on the same figure shows the theoretical estimate (\ref{eq:pos_90deg}) for the top edge vortex line.
It passes through the vorticity core, which means that $z_1$
calculated using the line vortex model is a reasonable prediction for the apparent position
of the vortex.
Note that, even in 2D, the position of the point vortex does not exactly match the position of maximum vorticity \cite[see][]{Wang_Eldredge_2013_tcfd}.

Figure~\ref{fig:cmp_re500}(\textit{b})
shows the normalized edge vortex circulation estimated by integration of the vorticity
over the selected windows. CFD results obtained with two different discretization
grids are shown: the original grid with 4.61 million cells and a refined grid with 9.96 million cells.
The difference between these two results is less than 0.2\% for all $r/c<5$,
and only becomes noticeable near the tip where the wing tip vortex enters in the integration domain.
The theoretical estimate for $\Gamma_1$ (\ref{eq:circ_90deg}) is in a good agreement
with the CFD results, with the difference being less than 17\% for all $r/c<5$.

\begin{figure}
\begin{center}
\includegraphics[scale=0.9]{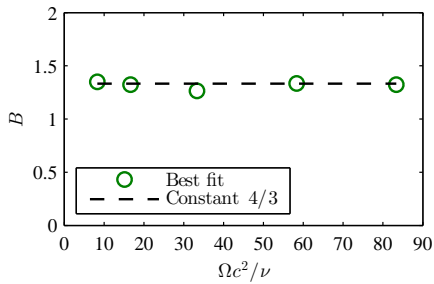}
\caption{\label{fig:power_vs_re} Optimal values of the power law exponent in $\Gamma_\Sigma^{Fit}$ that best-fit $\Gamma_\Sigma^{CFD}$ in
the least-mean-squares sense.}
\end{center}
\end{figure}

A remarkable property of the theoretical scaling law of $\Gamma_1$ with $r$ is that
the exponent in (\ref{eq:circ_90deg}) is independent of any parameters.
It is therefore important to determine the best power law for fitting the CFD results.
Therefore, we have carried out a two-parameter optimization of
\begin{equation}
\Gamma_\Sigma^{Fit} = A \left(\frac{r}{c}\right)^B
\end{equation}
and determined the values of $A$ and $B$ that minimize the root-mean-square deviation
with respect to $\Gamma_\Sigma^{CFD}$. The optimal values of $B$ are shown in figure~\ref{fig:power_vs_re}.
The mean value of $B$ over the considered range of $\Omega c^2/\nu$ is 1.32, which only differs by 1\%
from the theoretical estimate $4/3$ for the growth rate of $\Gamma_1$ with $r$.

\begin{figure}
\begin{center}
\includegraphics[scale=0.9]{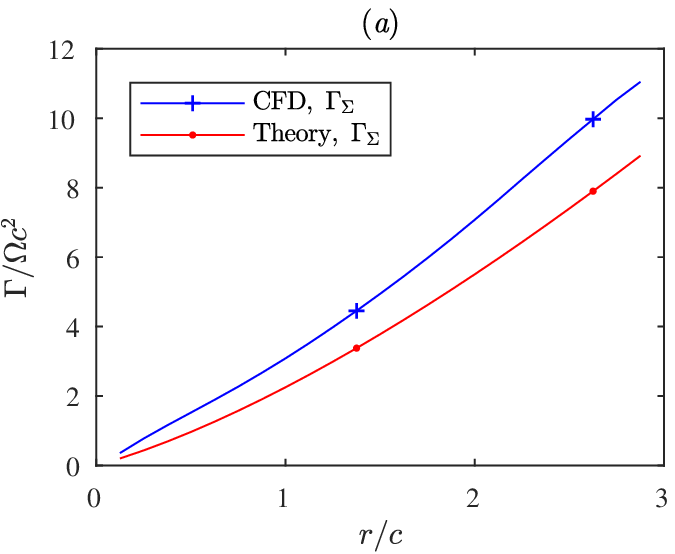}\quad
\includegraphics[scale=0.9]{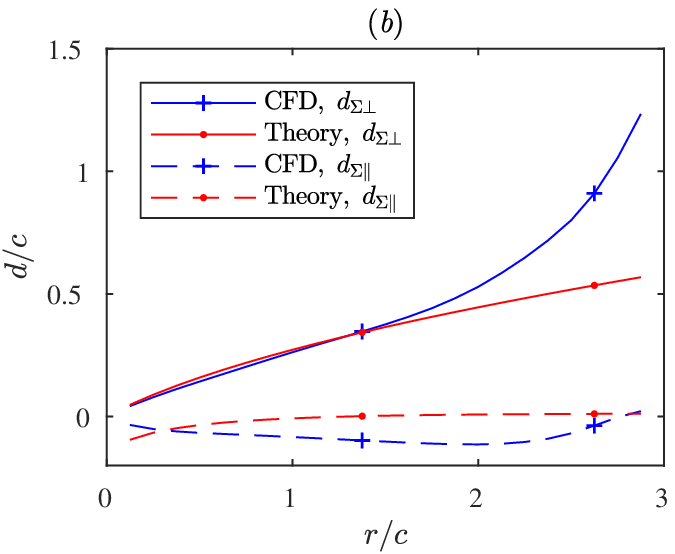}
\caption{\label{fig:gam_dist_ar3} A plate with the aspect ratio equal to 3. Comparison between the theoretical estimates and the CFD results
for (\textit{a}) the circulation $\Gamma_\Sigma$ over a cylinder surface of radius $r$, and
(\textit{b}) the distances between the edge and the vorticity centroids in
the directions parallel and perpendicular to the plate, $d_\parallel$ and $d_\perp$, respectively. All quantities are normalized.}
\end{center}
\end{figure}

The CFD data presented above is for a wing with the aspect ratio 6.
The wing length does not enter in our theoretical estimate,
but in the numerical simulation there may be some wing tip effects when the aspect ratio is small.
We have carried out an additional numerical simulation of a wing with the chord length $2$~mm, i.e.,
twice as wide as the original plate. The angular velocity is equal to $\Omega=260$~s$^{-1}$.
Apart from that, all parameters are the same. In particular, the wing length is equal to $6$~mm.
The aspect ratio is therefore equal to 3. The root-based Reynolds number is equal to
$\Omega c^2 / \nu = 67$.
The comparison between the theoretical and the numerical results is shown in figure~\ref{fig:gam_dist_ar3}.
The wing tip effects are significant over the distal part of the plate between $r/c=2$ and $3$.
Importantly, the extent of that domain is similar to what we found in the case of aspect ratio 6.
Over the proximal half of the plate, the agreement between the CFD results and the theory is good.

\subsection{Estimates of the average spanwise vorticity transport coefficient}

The algebraic growth rate of $\Gamma_1$ as $r^{4/3}$ is fully defined by the line vortex model,
but the prefactor in (\ref{eq:circ_90deg}) contains a parameter $K_{sp}$ that determines
how fast the Lagrangian elements of the line vortex are transported spanwise.
We therefore refer to $K_{sp}$ as the spanwise vorticity transport coefficient.
The exact value of $K_{sp}$ in each case depends on the distribution of the radial vorticity and the spanwise velocity in the flow field.
Consequently, it depends on $Re$, for the reason that the structure of the edge vortex varies significantly with $Re$.
Let us first derive a quick theoretical estimate of the spanwise vorticity transport coefficient $K_{sp}$
suitable for the low end of the range of $Re$ considered in the previous section.
Let $V_r(r,\phi,y)$ be the radial velocity component in the cylindrical polar coordinates.
At the plate, $\phi=0$ and the radial direction is aligned with the spanwise direction.
The vorticity transport in the radial direction mainly takes place at those locations
where both the radial vorticity $\omega_r$ and the radial velocity $V_r$
are large enough. To quantify it, we introduce the vorticity-weighted average radial velocity
\begin{equation}
%\overline{V}_r(r,\Phi) = \frac{ \int_{0}^{\Phi} \left( \int_{0}^{Ly/2} V_r \omega_r \mathrm{d} y \right) r \mathrm{d} \phi}{\int_{0}^{\Phi} \left( \int_{0}^{L_y/2} \omega_r \mathrm{d} y \right) r \mathrm{d} \phi},
\overline{V}_r(r,\Phi) = \frac{ \int_{0}^{\Phi} \int_{0}^{Ly/2} V_r \omega_r r \mathrm{d} y \mathrm{d} \phi}{\int_{0}^{\Phi} \int_{0}^{L_y/2} \omega_r r \mathrm{d} y \mathrm{d} \phi}.
\label{eq:vsp_estimate}
\end{equation}
The parameter $\Phi \in [0, 360^\circ]$ controls the extent of azimuthal averaging.
%which is a function of $r$ only.
%Integration over the top half-cylinder is possible as long as we restrict our attention
%to $\alpha=90$~deg, in which case the flow is symmetric with respect to the horizontal plane.
Further, the CFD results suggest that $V_r$ is approximately linear in $r$
over the inner-central part of the plate.
%in the vicinity of the plate near its root.
We therefore propose an estimate for the spanwise vorticity transport coefficient,
\begin{equation}
\overline{K}_{sp} = \overline{V}_r(r,\Phi) / \Omega r,
\label{eq:ksp_estimate}
\end{equation}
which we subsequently evaluate at a representative location $r_{ref}$. % and integrated over the complete half-cylinder $\Phi = 2 \pi$.
The overbar is to remind that the estimate is based on space averaging.

Near the plate, the vorticity is confined in two shear layers
that start from the edges and propagate in the downstream direction.
Due to the viscous exchange of momentum, the thickness of these vorticity sheets
increases with the distance from the edges, and the peak vorticity magnitude decreases.
We therefore use the one-dimensional diffusion equation in an unbounded domain to describe the
evolution of the vorticity profiles with the angular distance $\phi$ from the plate.
After introducing the time $t$ required for the plate to travel
the angular distance $\phi$, the vorticity is approximated as
\begin{equation}
\omega_r(r,t,y) = \frac{\gamma(r)}{\sqrt{4 \pi \nu t}} ( e^{-\frac{(y-c/2)^2}{4 \nu t}} - e^{-\frac{(y+c/2)^2}{4 \nu t}} ), \quad \mathrm{where} \quad t=\frac{\phi}{\Omega},
\label{eq:vort_prof}
\end{equation}
which satisfies the diffusion equation with the diffusivity equal to $\nu$,
and the initial condition corresponding to delta distribution of the vorticity at the sharp edges.

The radial velocity is mainly driven by the centrifugal forces acting on the fluid
trapped in the recirculation bubble,
and it also decays with the distance away from the plate due to the action of viscosity.
We assume the initial condition for $V_r$ of the form
$V_{r}(r,0,y) = V_{r~max}(r) \left( 1 - 4 y^2 /c^2 \right)$,
where, according to \cite{Maxworthy_2007_jfm},
$V_{r~max}(r) = \sqrt{2} \Omega r$.
The solution of the one-dimensional diffusion equation that satisfies the
initial condition is
\begin{equation}
\begin{split}
V_r(r,t,y) = V_{r~max}(r) \left\{ \left( \frac{2}{c^2}(y^2+2 \nu t)-\frac{1}{2} \right)
\left( \erf{\frac{y-c/2}{\sqrt{4 \nu t}}} - \erf{\frac{y+c/2}{\sqrt{4 \nu t}}} \right) + \right. \\
\left. \frac{4}{c^2} \sqrt{\frac{\nu t}{\pi}} \left( (y+\frac{c}{2}) e^{-\frac{(y-c/2)^2}{4 \nu t}} - (y-\frac{c}{2}) e^{-\frac{(y+c/2)^2}{4 \nu t}} \right) \right\}.
\end{split}
\label{eq:vsp_prof}
\end{equation}

\begin{figure}
\begin{center}
\includegraphics[scale=0.9]{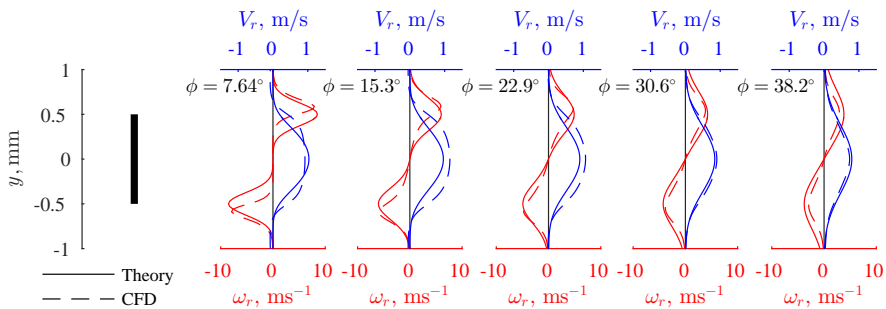}
\caption{\label{fig:om_vsp_prof} Line plots of the radial components of the vorticity $\omega_r$ (red) and the velocity $V_r$ (blue), sampled on vertical line segments between $y=-c/2$
and $y=c/2$, at a constant radial location $r_{ref}=3c$ and five different angular distances from the plate, $\phi=7.64^\circ n$, where $n=1,...,5$.
The solid lines show the analytical profiles (\ref{eq:vort_prof}) and (\ref{eq:vsp_prof}). The dash lines show the CFD results.
Vorticity magnitude is in milliseconds$^{-1}$.}
\end{center}
\end{figure}

Sample profiles of $\omega_r$ and $V_r$ are shown in figure~\ref{fig:om_vsp_prof}(\textit{a}) and compared with the CFD data.
They correspond to
a plate revolving with the angular velocity $\Omega = 260$~s$^{-1}$, i.e., $\Omega c^2 / \nu = 17$. The profiles are calculated at the radial location $r=3c$.
The parameter $\gamma(r)$ is set to $2.7$~m/s when evaluating (\ref{eq:vort_prof}), but it cancels out in the subsequent calculation of $\overline{K}_{sp}$.
The analytical profiles adequately describe the peaks of $\omega_r$ and $V_r$ as they flatten with the distance away from the plate.
It should be reminded, however, that the analytical profiles do not account for the dynamic coupling between $\omega_r$ and $V_r$ and for many three-dimensional effects that may
influence the rate of decay at larger $\phi$.
In the following, we use them to obtain a rough order of magnitude approximation to $\overline{K}_{sp}$ that does not rely on any data from the CFD.
On the other hand, to evaluate $\overline{K}_{sp}$ accurately, it is critical to account for the spatial distribution of $\omega_r$ and $V_r$ in all detail available from the CFD.

\begin{figure}
\begin{center}
\includegraphics[scale=0.9]{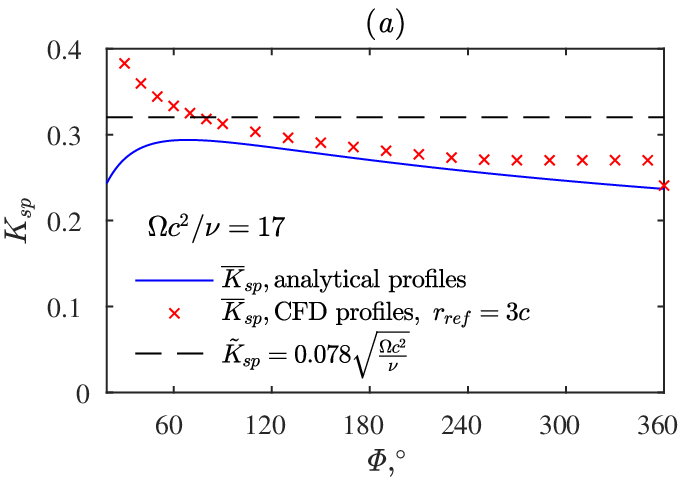}
\quad
\includegraphics[scale=0.9]{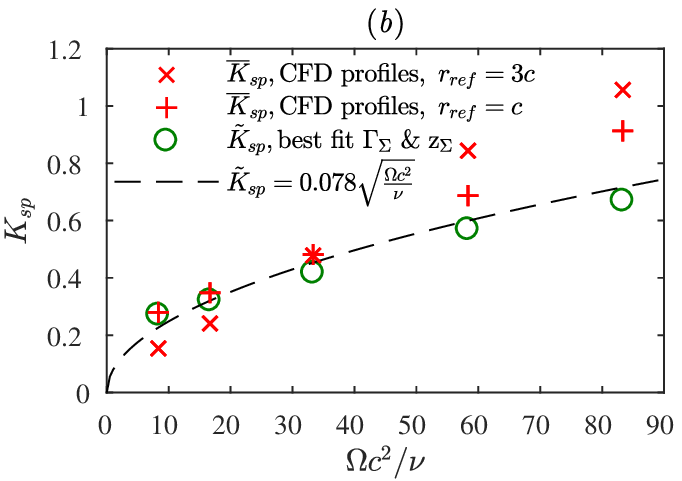}\\
\includegraphics[scale=0.9]{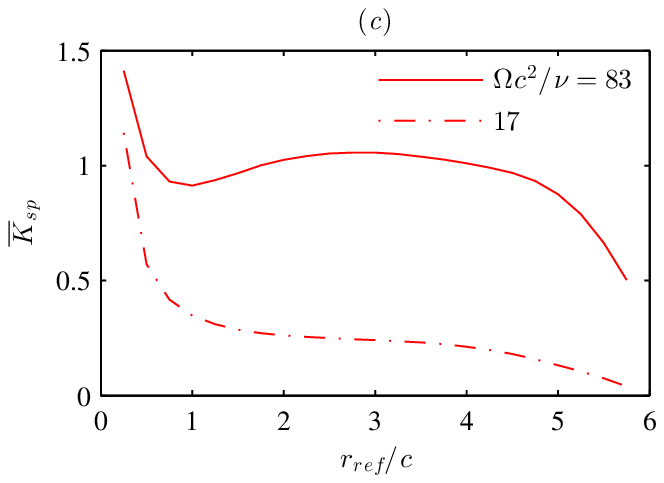}
\caption{\label{fig:ksp_vs_phi} Estimates of the spanwise vorticity transport coefficient $K_{sp}$: (\textit{a}) as a function of $\Phi$ at $\Omega c^2 / \nu = 17$,
$\Phi$ being the upper limit of integration in (\ref{eq:vsp_estimate});
(\textit{b}) as a function of the root-based Reynolds number $\Omega c^2 / \nu$, with $\Phi = 360^\circ$.
(\textit{c}) Spanwise distribution of the CFD-based estimate of the vorticity transport coefficient $\overline{K}_{sp}$, calculated
as given by (\ref{eq:vsp_estimate})-(\ref{eq:ksp_estimate}) with $\Phi = 360^\circ$.
%The horizontal axis is limited on the left by $\Phi_{min}=19^\circ$ that corresponds to one chord length distance travelled by the section of the plate located at $r_{ref}=3c$.
%Estimates with $\Phi < \Phi_{min}$ are inadequate because they do not include the recirculation bubble in the integration domain.
}
\end{center}
\end{figure}

After substituting (\ref{eq:vort_prof}) and (\ref{eq:vsp_prof}) into (\ref{eq:vsp_estimate}), performing numerical integration and substituting the result in (\ref{eq:ksp_estimate}), we obtain the
desired theoretical estimate for $\overline{K}_{sp}$. The result does not depend on $r$ because $V_{sp~max}(r)$ is linear in $r$.
Figure~\ref{fig:ksp_vs_phi}(\textit{a}) compares the values of $\overline{K}_{sp}$ calculated using the profiles (\ref{eq:vort_prof}) and (\ref{eq:vsp_prof}) with $\overline{K}_{sp}$ evaluated
using $\omega_r$ and $V_r$ from the CFD at $r_{ref}=3c$. The plots are shown in a range of $\Phi$ to explore the sensitivity %of the estimate (\ref{eq:ksp_estimate})
to this parameter. All values are within the interval between 0.23 and 0.4
when the linear distance from the plate is greater than $c$, i.e., $\Phi>19^\circ$.
The general trend is a slow decrease with $\Phi$. The CFD result saturates at $\Phi>240^\circ$ when the numerator and the denominator in $\overline{V}_r$
in (\ref{eq:vsp_estimate}) attain their finite maximum values. The sudden drop at $\Phi=360^\circ$ is explained by the inward spanwise velocity on the pressure side of the plate previously reported by \cite{Kolomenskiy_etal_2014_fdr}.

Figure~\ref{fig:ksp_vs_phi}(\textit{b}) displays $K_{sp}$ as a function of the root-based Reynolds number $\Omega c^2 / \nu$.
In addition to the estimate $\overline{K}_{sp}$ obtained by integration of $V_r$ and $\omega_r$ from the CFD,
the figure shows the values of $\tilde{K}_{sp}$ that best-fit the theoretical estimate %of $\Gamma_\Sigma^{Theory}$
to the CFD data %$\Gamma_\Sigma^{CFD}$
in the least-mean-squares sense, i.e.,
\begin{equation}
\sqrt{\left(\frac{\Gamma_\Sigma^{Theory}-\Gamma_\Sigma^{CFD}}{\Gamma_\Sigma^{CFD}}\right)^2+\left(\frac{|z_\Sigma^{Theory}-z_\Sigma^{CFD}|}{|z_\Sigma^{CFD}|}\right)^2} \to \min.
\label{eq:best_fit}
\end{equation}
A power law fit of those values leads to
the empirical formula (\ref{eq:ksp_estimate_Re}) that we used in the previous section.
The agreement between these different estimates is good except for large $\Omega c^2 / \nu$ and $r_{ref}$,
when the discrepancy of up to 50\% is caused by the vortex core structure becoming more complex and necessitating further investigation.
%Interestingly, $\overbar{K}_{sp}$ is greater than $\tilde{K}_{sp}$ in that case, which means that
%the spanwise flow is more than sufficient to explain
%and the vorticity annihilation effect \cite{Wojcik_Buchholz_2014_jfm}.
Apart from that, the estimate $\overline{K}_{sp}$ is consistent with $\tilde{K}_{sp}$ that matches the observed circulation and location of the vortex.
%The spanwise velocity is more than enough.

Two sample spanwise distributions of $\overline{K}_{sp}$,
obtained from the numerical simulations at $\Omega c^2 / \nu = 17$ and $83$, are shown in figure~\ref{fig:ksp_vs_phi}(\textit{c}).
In both cases, as postulated earlier, $\overline{K}_{sp}$ is roughly constant over the central part of the plate.
Variation only becomes large near the ends of the plate, i.e., $r_{ref}<1c$ or $r_{ref}>4.5c$.
Between these ends, the profile of $\overline{K}_{sp}$ depends on the Reynolds number: $\overline{K}_{sp}$ is monotonically decreasing
when the Reynolds number is small, but it has a local maximum when the Reynolds number is large.
Despite this small variability, the values sampled at $r_{ref}=1$ and $3$, used in figure~\ref{fig:ksp_vs_phi}(\textit{b}), are representative
of the average vorticity transport coefficient $\overline{K}_{sp}$ over the inner-central part of the plate
that we need for the edge vortex circulation and position estimates.

Let us conclude this section with a comment on the physical mechanisms
that drive the spanwise flow. This question has been extensively studied in the past research, and
several different mechanisms have been proposed.
Our objective is not to
describe all factors that may have certain influence on the spanwise velocity $V_r$,
but to quantify
the role of $V_r$ in the vorticity dynamics.
Our theoretical estimate (\ref{eq:vsp_prof})
is based on the model proposed by \cite{Maxworthy_2007_jfm}, who postulated that the centrifugal force and the outwards pressure gradient
in the conical vortex core are the two equally important drivers of $V_r$.
Other effects, such as the Coriolis acceleration and the wing-tip vortex induced velocity,
that are not accounted for in our model, are likely to have less influence on $K_{sp}$
compared with the two main effects postulated above.
For instance, the CFD computations by \cite{Garmann_2014_jfm} with the centrifugal term eliminated from the Navier--Stokes equations
show a dramatic decrease of the outwards spanwise velocity over the plate.
Even though the peak outwards spanwise velocity in the vortex core is positive and may be an
order of magnitude greater than the average \cite[][]{Garmann_2014_jfm,Limacher_etal_2016_jfm},
it is the average velocity that apparently matters for $K_{sp}$ and for the edge vortex dynamics,
as we infer from the overall good agreement between $\overline{K}_{sp}$ and $\tilde{K}_{sp}$ for the
conditions examined.
%The agreement between the theoretical results and the numerical simulations in the present work supports Maxworthy's %standpoint.

\subsection{Time evolution of the edge vortices}\label{sec:time_evolution}

The solution derived in \S\ref{sec:steady_theory} is steady.
However, the wing rotation starts from rest in our numerical simulations, as in many practical situations \cite[such as the
experiments by][using rectangular wings operating at $45^\circ$ angle of attack and $Re$ of order several thousand]{Carr_etal_2015_jfm}.
In addition to that, the flow may become unsteady due to hydrodynamic instabilities.
It is therefore important to consider the time evolution of the edge vortex.

\begin{figure}
\begin{center}
\includegraphics[scale=0.9]{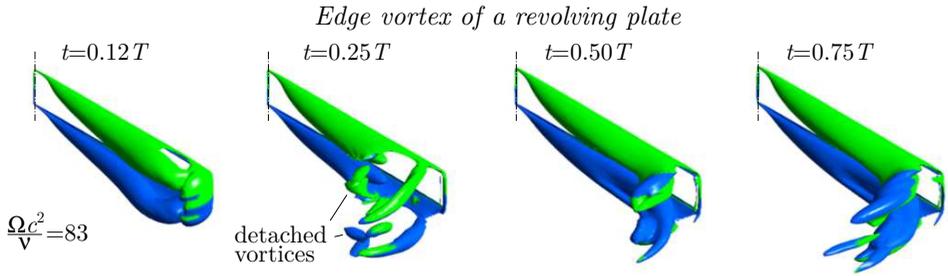}
\caption{\label{fig:iso_transient}
Instantaneous iso-surfaces of $\lambda_2 = -10^8$~s$^{-2}$, colored according to the sign of the spanwise vorticity component,
at four different time instants after startup.
The root-based Reynolds number is equal to $\Omega c^2 / \nu = 83$.
}
\end{center}
\end{figure}

%\begin{figure}
%\begin{center}
%\includegraphics[scale=0.9]{drawing_iso_all.eps}
%\caption{\label{fig:iso_transient}
%Instantaneous iso-surfaces of the $\lambda_2$-criterion at four different time instants
%after the startup,
%for two different values of the root-based Reynolds number:
%(\textit{a})  $\Omega c^2 / \nu = 17$ and (\textit{b}) $\Omega c^2 / \nu = 83$.
%The iso-surfaces correspond to $\lambda_2 = -5 \cdot 10^5$~s$^{-2}$ and
%$-10^8$~s$^{-2}$, respectively.
%They are colored according to the sign of the spanwise vorticity component.
%Images (\textit{a}) and (\textit{b}) are not on the same scale.
%}
%\end{center}
%\end{figure}

\begin{figure}
\begin{center}
\includegraphics[scale=0.9]{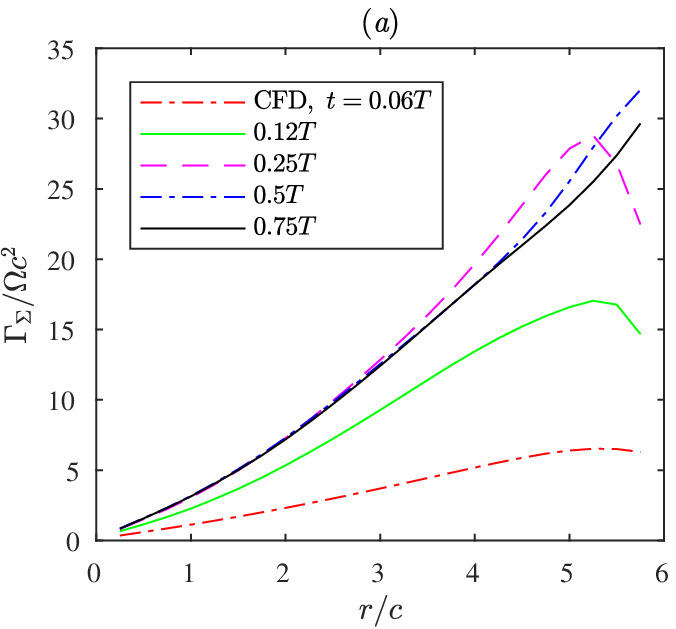}
\quad
\includegraphics[scale=0.9]{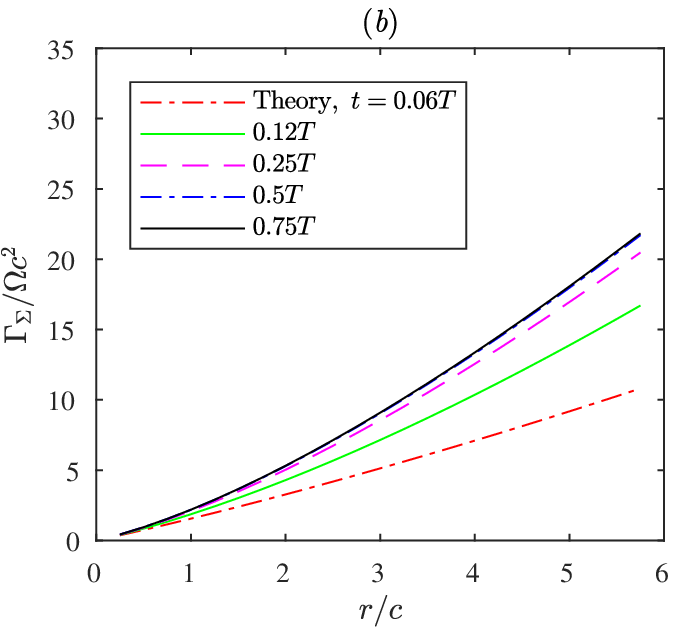}
\caption{\label{fig:circ_transient} Time evolution of the circulation in the case $\Omega c^2 / \nu = 83$.
(\textit{a}) Values obtained from (\ref{eq:circulation_cfd_90deg}) using the CFD data;
(\textit{b}) Theoretical values of $\Gamma_\Sigma$ as given by (\ref{eq:circ_90deg_transient}).
}
\end{center}
\end{figure}

Let us only discuss the largest Reynolds number case, $\Omega c^2 / \nu = 83$,
which illustrates well different kinds of unsteady effects. The aspect ratio of the plate is equal to 6.
We select the time instants
at $0.12T$, $0.25T$, $0.5T$ and $0.75T$ for the flow visualization, where $T = 2 \pi / \Omega = 4.8332 \cdot 10^{-3}$~s.
Time development of the vortex structure is illustrated by iso-surfaces of the $\lambda_2$-criterion in figure~\ref{fig:iso_transient}.
In addition, we plot the normalized circulation as a function of the normalized spanwise distance in figure~\ref{fig:circ_transient}(\textit{a}).
The vortices over the proximal part of the plate, $r<3c$, reach steady state by the time $t=0.25T$.
At the same time instant one can see a symmetric pair of counter-rotating vortices shed from the
distal part of the plate, $r>4.5c$. Later, the flow becomes nominally steady over $r<4.5c$,
but the wing-tip vortex is unsteady and small-scale eddies develop at this large Reynolds number.

Let us now amend our analysis to account for the gradual built-up of the edge vortex after the beginning of rotation.
Let $t$ be physical time with $t=0$ at the startup.
We extend the time profile of the plate angular velocity to negative $t$ as
\begin{equation}
\frac{\mathrm{d}\varphi_{plate}}{\mathrm{d}t} = \left\{
\begin{aligned}
& 0,  \quad t < 0, \\
& \frac{\Omega}{2} ( 1 - \cos \frac{\pi t}{t_{ac}} ),  \quad 0 \le t < t_{ac}, \\
& \Omega,  \quad t \ge t_{ac}.
\end{aligned}
\right.
\label{eq:kine_ext}
\end{equation}
Negative $t$ is the time before startup, when the plate and the surrounding fluid are at rest.
Large positive $t>t_{ac}$ is when the plate revolves steadily. Note that, even though our solution is defined for any arbitrary large $t$,
we are only interested in $t<T$ before the plate encounters its own wake from the previous revolution.

In the following analysis, the main difference with respect to the steady case is that now we track vortex particles
over a physical time interval from the startup until a set time instant.
The radial position $r$ of a tracer satisfies the evolution equation
\begin{equation}
\frac{\mathrm{d}r}{\mathrm{d}t} = K_{sp} \frac{\mathrm{d}\varphi_{plate}}{\mathrm{d}t} r.
\label{eq:lagr_traj_transient}
\end{equation}
Hence, the radial position of the tracer with the initial condition $r(0)=r_0$ can be written as
\begin{equation}
r = r_0 e^{K_{sp} \hat{t}_{ac} g(s) /2},
\label{eq:rad_pos_transient1}
\end{equation}
where
\begin{equation}
\hat{t}_{ac} = \frac{\Omega t_{ac}}{\pi} = 0.167, \quad
s = \frac{\Omega t}{\hat{t}_{ac}}, \quad \mathrm{and} \quad
g(s) = \left\{
\begin{aligned}
& 0,  \quad s < 0, \\
& s - \sin{s},  \quad 0 \le s < \pi, \\
& 2 s - \pi,  \quad s \ge \pi.
\end{aligned}
\right.
\label{eq:rad_pos_transient2}
\end{equation}

From the definition of $\tilde{\tau}$ (\ref{eq:tau_def}) we obtain
\begin{equation}
\tilde{\tau} = \frac{\Omega r_0^2 \hat{t}_{ac}}{4}
\int_{-\infty}^s g_s^2(s') e^{K_{sp} \hat{t}_{ac} g(s')} \mathrm{d} s',
\label{eq:tau_tilde_transient1}
\end{equation}
where the subscript $\cdot_s$ stands for the derivative.
Integration by parts yields
\begin{equation}
\tilde{\tau} = \frac{\Omega r_0^2}{4 K_{sp}} \left\{
\left. g_s(s') e^{K_{sp} \hat{t}_{ac} g(s')} \right|_{-\infty}^s -
\int_{-\infty}^s g_{ss}(s') e^{K_{sp} \hat{t}_{ac} g(s')} \mathrm{d} s'
\right\}
\label{eq:tau_tilde_transient2}
\end{equation}
We use a Taylor series approximation $e^x \approx 1+x$ for the exponential under the integral sign,
and express $r_0$ in terms of $r$ using (\ref{eq:rad_pos_transient1}). We thus obtain
\begin{equation}
\frac{\tilde{\tau}}{\Omega c^2} = \frac{1}{4 K_{sp}} \left(\frac{r}{c}\right)^2 \left\{
g_s(s) \left(1-e^{-K_{sp} \hat{t}_{ac} g(s)}\right) -
K_{sp} \hat{t}_{ac} f(s) e^{-K_{sp} \hat{t}_{ac} g(s)}
\right\},
\label{eq:tau_tilde_transient3}
\end{equation}
where
\begin{equation}
f(s) = \left\{
\begin{aligned}
& 0,  \quad s < 0, \\
& \frac{\sin{2s}}{4} + \sin{s} - s \sin{s} - \frac{s}{2},  \quad 0 \le s < \pi, \\
& \pi / 2,  \quad s \ge \pi.
\end{aligned}
\right.
\label{eq:h_of_s}
\end{equation}
The rest of the derivation is similar to the steady case. We finally obtain the position of the vortex
\begin{equation}
\frac{z_1}{c} = \frac{1}{2^{7/3}K_{sp}^{2/3}}
\left( \frac{r}{c} \right)^{2/3}
\left\{
1 - \left( 1 + K_{sp} \hat{t}_{ac} \frac{f(s)}{g_s(s)} \right) e^{-K_{sp} \hat{t}_{ac} g(s)}
\right\}^{2/3}
\label{eq:pos_90deg_transient}
\end{equation}
and its circulation
\begin{equation}
\frac{\Gamma_1}{\Omega c^2} = \frac{\pi}{(4K_{sp})^{1/3}} \left( \frac{r}{c} \right)^{4/3}
\frac{g_s(s)}{2} \left\{
1 - \left( 1 + K_{sp} \hat{t}_{ac} \frac{f(s)}{g_s(s)} \right) e^{-K_{sp} \hat{t}_{ac} g(s)}
\right\}^{1/3}.
\label{eq:circ_90deg_transient}
\end{equation}
The half-plate circulation $\Gamma_{half-plate}$ is calculated with the same formula as in
the steady case, see Appendix~\ref{sec:half-plate}, but using the time-dependent $\Gamma_1$ (\ref{eq:circ_90deg_transient}).

The sum circulation $\Gamma_{\Sigma}^{Theory} = \Gamma_1 + \Gamma_{half-plate}$ is shown in figure~\ref{fig:circ_transient}(\textit{b}),
for the same values of the aspect ratio and the Reynolds number as in the numerical simulation, and using
$K_{sp}$ as given by (\ref{eq:ksp_estimate_Re}).
The trend of $\Gamma_\Sigma$ increasing in time until it saturates is similar to what we observe in the numerical simulation,
but the theory predicts slightly smaller growth, and it does not account for the overshoot at $t=0.25T$ and $0.5T$ near the tip of the plate.
For small $t$, the vortex circulation $\Gamma_1$ is small, and
the largest contribution to $\Gamma_{\Sigma}^{Theory}$ is from the linear term $\sqrt{2} r / c$ in the half-plate bound circulation $\Gamma_{half-plate}$ (\ref{eq:circ_half_plate}).
As $t$ becomes large, the $r^{4/3}$ power law becomes dominant.
Similar trends were found in the
experiments by \cite{Carr_etal_2015_jfm}.

%The unsteady effects near the wing tip are, of course, beyond the scope of the theory.

%\subsection{\textcolor{red}{Edge vortices of a low aspect ratio plate}}\label{sec:ar3}

\section{Conclusions and perspectives}\label{sec:conclusions}

We have derived closed-form expressions for the edge vortex circulation $\Gamma_1$ and its position $z_1$, (\ref{eq:circ_90deg}) and (\ref{eq:pos_90deg}), respectively,
of a revolving plate at $90^\circ$ angle of attack.
The model only contains one free parameter, the spanwise vorticity transport coefficient $K_{sp}$.
For the latter, we have proposed a crude theoretical estimate (\ref{eq:ksp_estimate}) and a practical fit (\ref{eq:ksp_estimate_Re}) that minimizes the error of the circulation $\Gamma_\Sigma$.
The theoretical estimates of $\Gamma_\Sigma$ and $z_\Sigma$ are in a good agreement with the numerical solution of the Navier--Stokes equations in the root-based Reynolds number range $\Omega c^2 / \nu$ from 8 to 83.
Remarkably, the growth rate of $\Gamma_1$ as $r^{4/3}$ is independent of any parameters.
The vorticity production at the edge
and its three-dimensional transport are therefore sufficient to describe the edge vortex circulation, to the leading order.
Our model is not intended to explain the mechanisms that drive the spanwise flow, but the values of
$K_{sp}$ that we obtain are consistent with the theory by \cite{Maxworthy_2007_jfm}.

The flow considered in our study is similar to the LEV on a wing that operates at any large angle of attack.
Generalization of (\ref{eq:circ_90deg}) and (\ref{eq:pos_90deg}) appears feasible, but special care should be taken of the downwash which is not present in the current model,
which may require numerical solution of the Brown--Michael equation (\ref{eq:bm_motion}) and is therefore beyond the scope of this paper.
Likewise, the effect of non-zero distance between the wing root and the axis of rotation \cite[also known as petiolation, see][]{Phillips_2017_if}
may lend itself to modelling using the same vortex method, with special care taken of the flow near the wing root.
Finally, we emphasize that the mechanisms of stable attachment of LEVs are not well understood yet. %, and are often taken for granted \cite[see, e.g.,][]{Limacher_etal_2016_jfm}.
The success of the Brown--Michael vortex model to describe the edge vortex of a revolving plate, confirmed in the present study, opens a new perspective to analyze the stability of the leading-trailing vortex pair and the transition to periodic vortex shedding, using methods similar to those developed by \cite{Michelin_Smith_2009_tcfd}.

%Acknowledgements should be included at the end of the paper, before the References section or any appendicies, and should be a separate paragraph without a heading. Several anonymous individuals are thanked for contributions to these instructions.
%The authors thank Jean-Yves Andro, Jeff Eldredge and Keith Moffatt for fruitful discussions that have led to this study.

The authors thank Jean-Yves Andro and Keith Moffatt for many enlightening discussions that ultimately led to this study, and
Jeff Eldredge for his useful comments during the Thirteenth International Conference on Flow Dynamics.
DK gratefully acknowledges the financial support from the JSPS (Japan Society for the Promotion of Science)
Postdoctoral Fellowship, JSPS KAKENHI No. 15F15061.
DC was partly supported by a JASSO Honors Scholarship.
HL was partly supported by the JSPS KAKENHI No. 24120007 for Scientific
Research on Innovative Areas.
This work is dedicated in memory of Tony Maxworthy.

\appendix

\section{Error of the local point vortex approximation}\label{sec:error_local_approx}

\begin{figure}
\begin{center}
\includegraphics[scale=0.9]{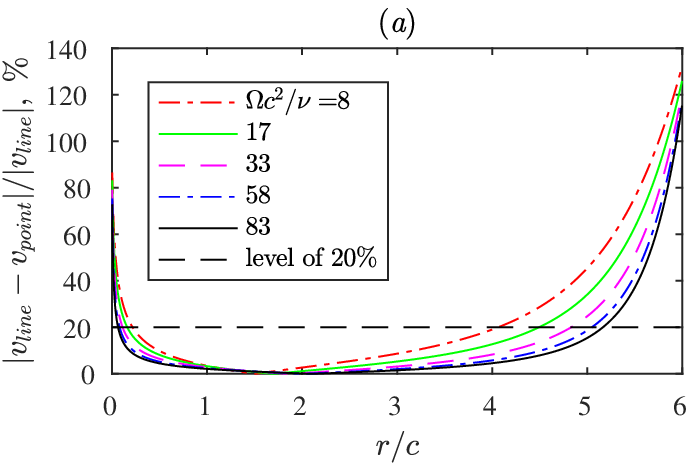}\quad
\includegraphics[scale=0.9]{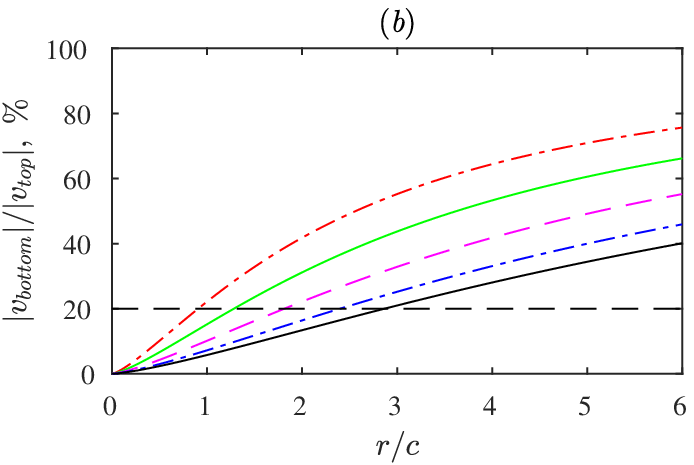}
\caption{\label{fig:line_approx} (\textit{a}) Relative difference between the induced velocity of a line vortex of variable strength and
a two-dimensional point vortex. (\textit{b}) Vertical velocity magnitude due to the bottom-edge vortex
relative to the velocity magnitude due to the top-edge vortex probed at the top edge.}
\end{center}
\end{figure}

The rightmost term in (\ref{eq:complex_pot}) is
the complex potential of a point vortex and its mirror image.
A point vortex is a two-dimensional approximation for a straight line
vortex in the three-dimensional flow that has constant circulation.
However, in our three-dimensional model, the circulation varies as $r^{4/3}$.
Therefore, the Kutta condition is not exactly satisfied.
With the shape of the vortex line and its circulation given by (\ref{eq:pos_90deg}) and (\ref{eq:circ_90deg}), respectively,
it is straightforward to use the Biot--Savart formula to compute the induced velocity at the edge of the plate.
In figure~\ref{fig:line_approx}(\textit{a}), it is compared with the induced velocity
in the local two-dimensional approximation. The relative difference is less than 20\%
in the range of $r/c$ between 0.3 and 4 in the examples considered in this paper.

A more significant error is to neglect the influence of the vortex generated by the bottom edge of the plate.
If the vertical velocity component induced by the top edge vortex is $v_{top}=\Gamma_1 / 2 \pi z_1$ (the imaginary part of $z_1$ in (\ref{eq:pos_90deg}) is zero),
then the vertical velocity component induced by the bottom edge vortex at the same point is $v_{bottom}=\Gamma_1 z_1 / 2 \pi (c^2 + z_1^2)$.
The ratio between the magnitudes of $v_{bottom}$ and $v_{top}$ is shown in figure~\ref{fig:line_approx}(\textit{b}).
For the largest Reynolds number, the ratio is of about 40\% at most, it is less than 20\% over the proximal half of the wing,
and 21\% on average over the span.
For the lowest Reynolds number, it is 50\% on average over the span.
This effect may explain larger discrepancy in the position of the vortex found in the comparison with the CFD results at low Reynolds numbers.

When the circulation of the radial vortex line varies over its length,
longitudinal vortices are produced such that the vortex system satisfies
the Helmholtz theorems. In particular, this effect explains the wing tip vortices.
The strength of the longitudinal vortices is related to the rate of change of the
edge vortex circulation with $r$, therefore,
their effect is likely to be of the same order of magnitude as
that of the non-uniform distribution of the circulation.
Detailed analysis of the three-dimensional wake is beyond the scope of this paper.
Note that the original model developed by \cite{Brown_Michael_1954_aiaa}
also applied the two-dimensional approximation to solve a three-dimensional problem, which was the LEV of a delta wing in that case.

\section{Bound circulation of the plate}\label{sec:half-plate}

The bound circulation corresponds to the vorticity contained in the boundary layers of the plate.
Let us calculate the circulation $\Gamma_{b}(\delta,\tau)$ along
a contour in the physical plane that begins at the pressure surface at a distance $\delta$ from the edge,
wraps around the edge but not the point vortex in the fluid domain, and ends at the suction surface at the same
distance $\delta$ from the edge. The beginning and the end points of the contour are, respectively,
$z_- = \lim_{\epsilon \to 0}(-\epsilon-\mathrm{i}\delta)$ and
$z_+ = \lim_{\epsilon \to 0}(\epsilon-\mathrm{i}\delta)$, where $\epsilon \in \mathbb{R}^+$.
The direction is consistent with our sign convention for the circulation.
Knowing the complex potential (\ref{eq:complex_pot}), the bound circulation is equal to
\begin{equation}
\Gamma_{b}(\delta,\tau) = \Re\left\{W\left(\zeta(z_+),\tau\right)-W\left(\zeta(z_-),\tau\right)\right\}.
\end{equation}
Noting that $\zeta(z_+)=\sqrt{c \delta}$ and $\zeta(z_-)=-\sqrt{c \delta}$,
we obtain
\begin{equation}
\frac{\Gamma_{b}(\delta,\tau)}{\Omega c^2} = 2\frac{r}{c}\sqrt{\frac{\delta}{c}} - \frac{\Gamma_1}{\Omega c^2} \frac{1}{\pi}
\left\{ \mathrm{Arg}\left( \sqrt{c \delta} - \zeta_1 \right) - \mathrm{Arg}\left( -\sqrt{c \delta} - \zeta_1 \right) + \pi n \right\}, ~~ n \in \mathbb{Z}.
\end{equation}
The value of $n$ is determined by requiring $\Gamma_{b}$ to be continuous with respect to $\delta$ and vanishing as $\delta \to 0$. After expressing $\mathrm{Arg}$ in terms of trigonometric functions
and using the fact that $z_1$ is real, we find
\begin{equation}
\frac{\Gamma_{b}(\delta,\tau)}{\Omega c^2} = 2 \frac{r}{c} \sqrt{\frac{\delta}{c}} - \frac{\Gamma_1}{\Omega c^2}
\left\{ \frac{1}{2} + \frac{1}{\pi} \arctan{\frac{\delta-z_1}{\sqrt{2 \delta z_1}}} \right\}.
\label{eq:gammab}
\end{equation}
In this work, we use (\ref{eq:gammab}) evaluated at $\delta=c/2$ as an approximation to
the bound circulation of the upper half of a finite plate of chord $c$, i.e., $\Gamma_{half-plate} \approx \Gamma_{b}(c/2,\tau)$. This is consistent with the original semi-infinite plate assumption of this study.
More accurate account of the bound vorticity distribution over a finite plate is possible, but in general it requires
numerical integration. Though it may change the result quantitatively by as much as 41\% (in the limiting case of $\Gamma_1=0$)
comparing with the above estimate at $\delta=c/2$, the qualitative trends are not changed. Since $\Gamma_{half-plate}$ is, in practice,
small compared with $\Gamma_1$, the approximation is adequate.

The position of the half-plate bound vorticity center is defined as
\begin{equation}
z_{half-plate}=-\mathrm{i} \delta_{half-plate},
\label{eq:zhp_def}
\end{equation}
where
$\delta_{half-plate}$ is the distance from the edge of the plate to the half-plate bound vorticity center,
\begin{equation}
\delta_{half-plate} = \frac{1}{\Gamma_{half-plate}} \int_0^{c/2}{\delta \frac{\mathrm{d} \Gamma_b}{\mathrm{d} \delta} \mathrm{d} \delta}.
\label{eq:dhp_def}
\end{equation}
Taking the derivative of (\ref{eq:gammab}), we obtain
\begin{equation}
\frac{\mathrm{d} \Gamma_b}{\mathrm{d} \delta} = \Omega r \sqrt{\frac{c}{\delta}} - \frac{\Gamma_1}{\pi}
\sqrt{\frac{z_1}{2 \delta}} \frac{z_1+\delta}{z_1^2+\delta^2}.
\label{eq:gammahp_derivative}
\end{equation}
From (\ref{eq:zhp_def}), (\ref{eq:dhp_def}) and (\ref{eq:gammahp_derivative}),
dividing the result by $c$, we obtain
the normalized position of the half-plate bound vorticity center,
\begin{equation}
\frac{z_{half-plate}}{c} = -\mathrm{i} \frac{ \frac{1}{3\sqrt{2}}\frac{r}{c} - \frac{\Gamma_1}{\Omega c^2}\frac{z_1}{\pi c} \left( \frac{1}{2} \log{
\frac{ \frac{1}{2}+\frac{z_1}{c}-\sqrt{\frac{z_1}{c}} }{ \frac{1}{2}+\frac{z_1}{c}+\sqrt{\frac{z_1}{c}} }
} + \sqrt{\frac{c}{z_1}} \right) }{\sqrt{2} \frac{r}{c} - \frac{\Gamma_1}{\Omega c^2} \left( \frac{1}{2} + \frac{1}{\pi} \arctan{\frac{1-2z_1/c}{2\sqrt{z_1/c}}} \right)}.
\label{eq:z_half-plate}
\end{equation}

\bibliographystyle{jfm}
\bibliography{lev_jfm}

\begin{thebibliography}{20}
\expandafter\ifx\csname natexlab\endcsname\relax\def\natexlab#1{#1}\fi
\def\au#1{#1} \def\ed#1{#1} \def\yr#1{#1}\def\at#1{#1}\def\jt#1{\textit{#1}}
  \def\bt#1{#1}\def\bvol#1{\textbf{#1}} \def\vol#1{#1} \def\pg#1{#1}
  \def\publ#1{#1}\def\arxiv#1{#1}\def\org#1{#1}\def\st#1{\textit{#1}}

\bibitem[Birch \& Dickinson(2001)]{Birch_Dickinson_2001_nature}
{\sc \au{Birch, J.~M.} \& \au{Dickinson, M.~H.}} \yr{2001}  \at{Spanwise flow
  and the attachment of the leading-edge vortex on insect wings}.  \jt{Nature}
  \bvol{412}~(6848),  \pg{729--733}.

\bibitem[Brown \& Michael(1954)]{Brown_Michael_1954_aiaa}
{\sc \au{Brown, C.~E.} \& \au{Michael, W.~H.}} \yr{1954}  \at{Effect of
  leading-edge separation on the lift of a delta wing}.  \jt{Journal of the
  Aeronautical Sciences}  \bvol{21}~(10),  \pg{690--694}.

\bibitem[Carr {\em et~al.\/}(2015)Carr, De{V}oria \&
  Ringuette]{Carr_etal_2015_jfm}
{\sc \au{Carr, Z.~R.}, \au{De{V}oria, A.~C.} \& \au{Ringuette, M.~J.}}
  \yr{2015}  \at{Aspect-ratio effects on rotating wings: circulation and
  forces}.  \jt{Journal of Fluid Mechanics}  \bvol{767},  \pg{497--525}.

\bibitem[Cortelezzi(1995)]{Cortelezzi_1995_pof}
{\sc \au{Cortelezzi, L.}} \yr{1995}  \at{On the unsteady separated flow past a
  semi-infinite plate: Exact solution of the {B}rown and {M}ichael model,
  scaling, and universality}.  \jt{Physics of Fluids}  \bvol{7}~(3),  \pg{526}.

\bibitem[Ellington {\em et~al.\/}(1996)Ellington, van~den Berg, Willmott \&
  Thomas]{Ellington_etal_1996_nature}
{\sc \au{Ellington, C.~P.}, \au{van~den Berg, C.}, \au{Willmott, A.~P.} \&
  \au{Thomas, A. L.~R.}} \yr{1996}  \at{Leading-edge vortices in insect
  flight}.  \jt{Nature}  \bvol{384}~(6610),  \pg{626--630}.

\bibitem[Garmann \& Visbal(2014)]{Garmann_2014_jfm}
{\sc \au{Garmann, D.~J.} \& \au{Visbal, M.~R.}} \yr{2014}  \at{Dynamics of
  revolving wings for various aspect ratios}.  \jt{Journal of Fluid Mechanics}
  \bvol{748},  \pg{932--956}.

\bibitem[Garmann {\em et~al.\/}(2013)Garmann, Visbal \&
  Orkwis]{Garmann_2013_pof}
{\sc \au{Garmann, D.~J.}, \au{Visbal, M.~R.} \& \au{Orkwis, P.~D.}} \yr{2013}
  \at{Three-dimensional flow structure and aerodynamic loading on a revolving
  wing}.  \jt{Physics of Fluids}  \bvol{25}~(3),  \pg{034101}.

\bibitem[Harbig {\em et~al.\/}(2013)Harbig, Sheridan \&
  Thompson]{Harbig_etal_2013_jfm}
{\sc \au{Harbig, R.~R.}, \au{Sheridan, J.} \& \au{Thompson, M.~C.}} \yr{2013}
  \at{Reynolds number and aspect ratio effects on the leading-edge vortex for
  rotating insect wing planforms}.  \jt{Journal of Fluid Mechanics}
  \bvol{717},  \pg{166--192}.

\bibitem[Kolomenskiy {\em et~al.\/}(2014)Kolomenskiy, Elimelech \&
  Schneider]{Kolomenskiy_etal_2014_fdr}
{\sc \au{Kolomenskiy, D.}, \au{Elimelech, Y.} \& \au{Schneider, K.}} \yr{2014}
  \at{Leading-edge vortex shedding from rotating wings}.  \jt{Fluid Dynamics
  Research}  \bvol{46},  \pg{031421}.

\bibitem[Kruyt {\em et~al.\/}(2015)Kruyt, van Heijst, Altshuler \&
  Lentink]{Kruyt_etal_2015_interface}
{\sc \au{Kruyt, J.~W.}, \au{van Heijst, G.~F.}, \au{Altshuler, D.~L.} \&
  \au{Lentink, D.}} \yr{2015}  \at{Power reduction and the radial limit of
  stall delay in revolving wings of different aspect ratio}.  \jt{Journal of
  The Royal Society Interface}  \bvol{12}~(105),  \pg{20150051}.

\bibitem[Lentink \& Dickinson(2009)]{Lentink_Dickinson_2009b_jeb}
{\sc \au{Lentink, D.} \& \au{Dickinson, M.~H.}} \yr{2009}  \at{Rotational
  accelerations stabilize leading edge vortices on revolving fly wings}.
  \jt{The Journal of experimental biology}  \bvol{212},  \pg{2705--2719}.

\bibitem[Limacher {\em et~al.\/}(2016)Limacher, Morton \&
  Wood]{Limacher_etal_2016_jfm}
{\sc \au{Limacher, E.}, \au{Morton, C.} \& \au{Wood, D.}} \yr{2016}  \at{On the
  trajectory of leading-edge vortices under the influence of coriolis
  acceleration}.  \jt{Journal of Fluid Mechanics}  \bvol{800},  \pg{R1}.

\bibitem[Liu {\em et~al.\/}(1998)Liu, Ellington, Kawachi, van~den Berg \&
  Willmott]{Liu_etal_1998_jeb}
{\sc \au{Liu, H.}, \au{Ellington, C.~P.}, \au{Kawachi, K.}, \au{van~den Berg,
  C.} \& \au{Willmott, A.~P.}} \yr{1998}  \at{A computational fluid dynamic
  study of hawkmoth hovering}.  \jt{Journal of Experimental Biology}
  \bvol{201}~(4),  \pg{461--477}.

\bibitem[Maxworthy(1979)]{Maxworthy_1979_jfm}
{\sc \au{Maxworthy, T.}} \yr{1979}  \at{Experiments on the {W}eis-{F}ogh
  mechanism of lift generation by insects in hovering flight. {P}art 1.
  {D}ynamics of the `fling'}.  \jt{Journal of Fluid Mechanics}  \bvol{93}~(1),
  \pg{47--63}.

\bibitem[Maxworthy(2007)]{Maxworthy_2007_jfm}
{\sc \au{Maxworthy, T.}} \yr{2007}  \at{The formation and maintenance of a
  leading-edge vortex during the forward motion of an animal wing}.
  \jt{Journal of Fluid Mechanics}  \bvol{587}~(2007),  \pg{471--475}.

\bibitem[Michelin \& Llewellyn~{S}mith(2009)]{Michelin_Smith_2009_tcfd}
{\sc \au{Michelin, S.} \& \au{Llewellyn~{S}mith, S.~G.}} \yr{2009}  \at{An
  unsteady point vortex method for coupled fluid--solid problems}.
  \jt{Theoretical and Computational Fluid Dynamics}  \bvol{23}~(2),
  \pg{127--153}.

\bibitem[Phillips {\em et~al.\/}(2017)Phillips, Knowles \&
  Bomphrey]{Phillips_2017_if}
{\sc \au{Phillips, N.}, \au{Knowles, K.} \& \au{Bomphrey, R.~J.}} \yr{2017}
  \at{Petiolate wings: effects on the leading-edge vortex in flapping flight}.
  \jt{Interface Focus}  \bvol{7}~(1),  \pg{20160084}.

\bibitem[Usherwood \& Ellington(2002)]{Usherwood_Ellington_2002b_jeb}
{\sc \au{Usherwood, J.~R.} \& \au{Ellington, C.~P.}} \yr{2002}  \at{The
  aerodynamics of revolving wings {I}{I}. {P}ropeller force coefficients from
  mayfly to quail}.  \jt{Journal of Experimental Biology}  \bvol{205}~(11),
  \pg{1565--1576}.

\bibitem[Wang \& Eldredge(2013)]{Wang_Eldredge_2013_tcfd}
{\sc \au{Wang, C.} \& \au{Eldredge, J.~D.}} \yr{2013}  \at{Low-order
  phenomenological modeling of leading-edge vortex formation}.  \jt{Theoretical
  and Computational Fluid Dynamics}  \bvol{27}~(5),  \pg{577--598}.

\bibitem[Wojcik \& Buchholz(2014)]{Wojcik_Buchholz_2014_jfm}
{\sc \au{Wojcik, C.~J.} \& \au{Buchholz, J. H.~J.}} \yr{2014}  \at{Vorticity
  transport in the leading-edge vortex on a rotating blade}.  \jt{Journal of
  Fluid Mechanics}  \bvol{743},  \pg{249--261}.

\end{thebibliography}

\end{document}